\documentclass[aps,prl,showpacs,twocolumn,footinbib,superscriptaddress,floatfix]{revtex4}

\usepackage{amsmath}
\usepackage{amssymb}
\usepackage{color}
\usepackage{epsfig}
\usepackage{fancyhdr}
\usepackage{graphicx}
\usepackage{latexsym}
\usepackage{tabularx}
\usepackage{times}

\newcommand{\Z}{\mathbb{Z}}

\newcommand{\sss}{\scriptscriptstyle}
\newcommand{\braket}[1]{\left\langle{#1}\right\rangle}

\newcommand{\av}[1]{\left[{#1}\right]_{\sss\rm av}}
\newcommand{\aT}[1]{\braket{#1}_{\sss\rm T}}

\newcommand{\theresult}{0.055(2)}
\newcommand{\theresultpercent}{5.5(2)\%}
\newcommand{\thesample}{0.048}
\newcommand{\thesampleTc}{1.251(8)}
\newcommand{\thesamplepercent}{4.8\%}

\begin{document}

\title{Optimal Error Correction in Topological Subsystem Codes}

\author{Ruben S.~Andrist}
\affiliation{Theoretische Physik, ETH Zurich, CH-8093 Zurich,
Switzerland}

\author{H.~Bombin}
\affiliation{Perimeter Institute for Theoretical Physics, Waterloo,
Ontario N2L 2Y5, Canada}

\author{Helmut G.~Katzgraber}
\affiliation {Department of Physics and Astronomy, Texas A\&M
University, College Station, Texas 77843-4242, USA}
\affiliation {Theoretische Physik, ETH Zurich, CH-8093 Zurich,
Switzerland}

\author{M.~A.~Martin-Delgado}
\affiliation{Departamento de F{\'i}sica Te{\'o}rica I, Universidad
Complutense, 28040 Madrid, Spain}



\begin{abstract}

A promising approach to overcome decoherence in quantum computing
schemes is to perform active quantum error correction using topology.
Topological subsystem codes incorporate both the benefits of topological
and subsystem codes, allowing for error syndrome recovery with only
2-local measurements in a two-dimensional array of qubits.  We study the
error threshold for topological subsystem color codes under very general
external noise conditions.  By transforming the problem into a classical
disordered spin model, we estimate using Monte Carlo simulations that
topological subsystem codes have an optimal error tolerance of
$\theresultpercent$. This means there is ample space for improvement in
existing error-correcting algorithms that typically find a threshold of
approximately $2\%$.

\end{abstract}

\date{\today}
\maketitle

Quantum computing promises to fundamentally further the bounds of
computability, particularly in such fields as complexity theory and
cryptography, and, in particular, the simulation of chemical and
physical systems. Unfortunately, implementations of quantum computing
proposals require precise manipulations of quantum systems which are
highly susceptible to external noise. The technical feasibility of
any quantum computer design thus heavily relies on efficient quantum
error detection and recovery. This can be achieved, for example,
by redundantly encoding quantum information in a code subspace
of many physical qubits \cite{shor:95,steane:96,knill:97}. Such
a suitable subspace is defined in terms of stabilizer operators
\cite{gottesman:96,calderbank:96}---products of individual Pauli
operators---and their corresponding eigenvalues.

Because stabilizers need to be measured during the error
recovery procedure, geometric locality of the involved qubits is
essential for practicality. Topological error correcting codes
\cite{kitaev:03,bombin:06,bombin:07,bombin:10,bravyi:10,haah:11}
achieve this by arranging qubits on a topologically
nontrivial manifold with stabilizers acting only on
neighboring qubits. These codes promise a reliable approach
to quantum computing, because of their stability to errors
\cite{dennis:02,raussendorf:07,katzgraber:09c,barrett:10,duclos:10,wang:11,landahl:11}:
A sizable fraction of physical qubits needs to fail before the logical
information encoded in the system is lost beyond error correction.

To determine the error stability of topologically protected
quantum computing proposals it is customary to map the
error correction procedure onto the thermodynamic behavior of
a disordered classical (statistical-mechanical) spin system
\cite{dennis:02,katzgraber:09c,wang:03}. There is a fruitful synergy
between quantum computation and statistical mechanics:
On the one hand, the stability of quantum computing proposals can be
studied with the well-established machinery from statistical physics
of complex systems, and on the other hand, it also opens the door to exotic
applications of statistical models.

Unfortunately, there is one caveat: The stabilizers for surface codes
(such as the Kitaev code \cite{kitaev:03}) and topological color codes
\cite{bombin:06} involve multiple qubits---four in the case of the
Kitaev code, six or eight for color codes.  This immensely complicates
physical realizations.  However, in stabilizer subsystem codes
\cite{poulin:05,bacon:06} some of the encoded logical qubits are ``gauge
qubits'' where no information is encoded. This provides ancilla qubits
to absorb decoherence effects and, in particular, allows breaking up the
required measurements for error recovery into several individual
measurement that involve a smaller number of qubits
\cite{bacon:06,poulin:05}, e.g., two.  Hence, physical realizations
are more feasible at the price of requiring additional qubits. 
Note that extensions and variants have also been proposed
\cite{bravyi:11,crosswhite:11}.

\begin{figure}
\includegraphics[width=\columnwidth]{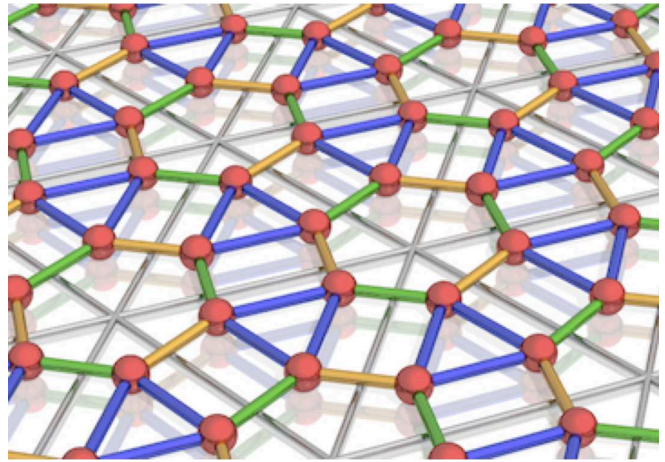}
\caption{(Color online)
Graphical representation of the qubit arrangement for topological
subsystem color codes on a regular triangular lattice. Each of the triangular
unit cells (large gray triangles) contains three physical qubits (red
balls).  The two-qubit gauge generators $\sigma^w\otimes\sigma^w$ are
shown in green ($w = x$), yellow ($w = y$) and blue ($w = z$). These are
the lines connecting the qubits (red balls).  They are arranged such
that each physical qubit has two generators of $z$ type, one of $x$ type
and one of $y$ type. See main text for details.
}
\label{fig:lattice}
\end{figure}

A true advantage is given by {\em topological} subsystem codes
\cite{bombin:10} which combine the robustness of topologically based
implementations with the simplicity of subsystem codes where only
measurements of neighboring qubits are required for recovery.  As in the
case of surface and color codes, the ideal error stability for
topological subsystem codes can be computed by mapping the error
recovery problem onto a classical statistical-mechanical Ising spin
system where the disorder corresponds to faulty physical qubits. Here, using
large-scale Monte Carlo simulations we compute the ideal error
correction threshold for topological subsystem color codes affected by
depolarizing noise. Our results show error correction is feasible up to
$\theresultpercent$ faulty physical qubits.  Remarkably, existing error
correcting algorithms only reach a threshold of approximately $2\%$
\cite{bombin:11,suchara:11}, leaving ample room for
improvement.

\paragraph*{Topological subsystem codes and mapping.---}
\label{sec:tsc}

A stabilizer subsystem code is defined by its gauge group $\mathcal G$.
Its elements are Pauli operators that, by definition, do not affect
encoded states. Namely, two states $\rho$ and $\rho'$ are equivalent if
$\rho=\sum_i g_i \rho g_i'$ with $g_i$ and $g_i'$ elements in the
algebra generated by $\mathcal G$.

\begin{figure}
\includegraphics[width=\columnwidth]{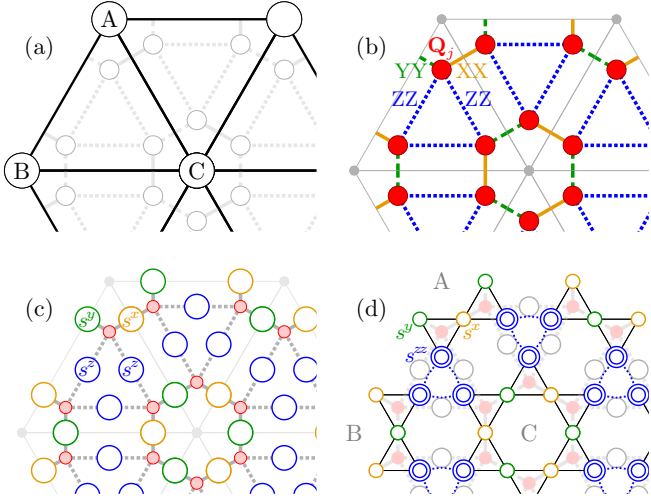}
\caption{(Color online)
(a) A regular triangular lattice satisfies the vertex three-colorability
requirement (indicated by A, B, C).  (b) To construct a
topological subsystem code, we place three qubits (red balls) inside
each of the triangular unit cells and connect them with
$\sigma^z\otimes\sigma^z$ gauge generators (dotted blue lines).  The
links between these triangles are assigned $\sigma^x\otimes\sigma^x$ and
$\sigma^y\otimes\sigma^y$ gauge generators (yellow and green 
solid lines, respectively).
(c) For the mapping, gauge generators represented by colored lines
in (b) are associated with Ising spins $s^{x,y,z}$ and the qubits with
interactions. (d) Introducing new Ising spin variables $s^{zz} = s^z
s'^z$ allows for the removal of local $\Z_2$ symmetries.
}
\label{fig:construction}
\end{figure}

Topological subsystem color codes \cite{bombin:10} are constructed
by starting from a two-dimensional lattice with triangular faces and
three-colorable vertices. Here we consider the triangular lattice shown
in Figs.~\ref{fig:lattice} and \ref{fig:construction}(a).  As indicated
in Figs.~\ref{fig:lattice} and \ref{fig:construction}(b), there are
three physical qubits per triangle and the gauge group has 2-local
generators $G_i$ of the form $\sigma^w\otimes\sigma^w$, where $w=x$,
$y$, and $z$.

Any family of topological codes shows a finite threshold for a given
local noise source. In other words, when the intensity of the noise is
below the threshold, we can correct errors with any desired accuracy at
the price of choosing a large enough code in the family. We are
interested in the error threshold of topological subsystem codes under
the effects of depolarizing noise, where each qubit is affected by a 
channel of the form
\begin{equation}
\mathcal{D}_p(\rho) = (1-p)\rho
	+\frac p 3 \sum_{w=x,y,z}
	\sigma^w\!\rho\sigma^w\,.
\label{eq:depol_channel}
\end{equation}
Here $\rho$ represents the density matrix describing the quantum state 
of the qubit and $p\in[0,1]$ its the probability for an error to occur.
The depolarizing channel plays a fundamental role in quantum information 
protocols where the effects of noise need to be considered, e.g., in 
quantum cryptography \cite{shor:00,kraus:05}, quantum distillation of 
entanglement \cite{bennett:96}, and quantum teleportation \cite{bowen:01}.

It is expected that there exists a threshold value $p=p_c$ such that in
the limit of large codes, for $p < p_c$ error correction succeeds with
probability 1 and for $p>p_c$ the result is entirely random. Remarkably,
for topological codes in general, one can relate $p_c$ to a phase
transition in a suitably-chosen classical disordered Ising spin model,
as we detail next.

To construct the related classical statistical-mechanical system, we
place an Ising spin $s_i=\pm1$ for each gauge generator $G_i$. Single
qubit Pauli operators $\sigma^w$ are mapped onto interaction terms
according to the generators $G_i$ with which they do not commute,
giving rise to a Hamiltonian of the general form
\begin{equation}
\mathcal{H}_\tau(s) := -J
	\sum_j
	\sum_{w=x,y,z}
	\tau_j^w \prod_i s_i^{g_{ij}^w}\,.
\label{eq:generalhamiltonian}
\end{equation}
Here $i$ enumerates all Ising spins and $j$ all physical qubit sites,
respectively.  For each spin $s_i$ the exponent $g_{ij}^w\in\{0,1\}$ is
0 [1] if $\sigma^w_j$ [anti]commutes with $G_i$. The signs of the
couplings $\tau_j^w=\pm 1$ are then quenched random variables
satisfying the constraint $\tau_j^x\tau_j^y\tau_j^z=1$. For each $j$,
they are all positive with probability $1-p$ and the other three
configurations have probability $p/3$ each.

In our specific case the Hamiltonian has the geometry depicted in
Fig.~\ref{fig:construction}(c) and thus takes the form
\begin{equation}
\mathcal{H} = -J\sum_j^n 
	(\tau_j^x s_j^y + \tau_j^y s_j^x) s_j^z\bar s_j^z
	+\tau_j^z s_j^x s_j^y\,,
\label{eq:hamiltonian}
\end{equation}
where $j$ enumerates qubit sites and spins are labeled, for each $j$, as
shown in Fig.~\ref{fig:construction}. Notice that $z$-labeled spins are
arranged in triangles, and that flipping each of these triads of spins
together does not change the energy of the system. Therefore, there is a
$\Z_2$ gauge symmetry. We fix the $\Z_2$ gauge symmetry and at the same
time simplify the Hamiltonian by introducing new Ising variables
$s_j^{zz}=s_j^z\bar s_j^z$. Notice that these spins are constrained: If
$j$, $k$, $l$ are three-qubit sites in a triangle,
$s_j^{zz}s_k^{zz}s_l^{zz}=1$. The simulated Hamiltonian therefore reads
\cite{comment:units}
\begin{equation}
\mathcal{H} = -J\sum_j^n 
	\tau_j^x s_j^xs_j^{zz}
	+\tau_j^y s_j^ys_j^{zz}
	+\tau_j^z s_j^xs_j^y\, .
\label{eq:hamiltonian2}
\end{equation}
Note that the Hamiltonian in Eq.~\eqref{eq:hamiltonian2} has no local
symmetries, but a global $\Z_2\times\Z_2$ symmetry. Indeed, we can color
spins according to their nearest colored vertex in the original 
lattice [Fig.~\ref{fig:construction}(a)], producing three sublattices A, 
B, and C.  Flipping the spins of two of these sublattices together 
leaves the energy invariant, giving rise to the indicated symmetry.

We are thus left with a random spin system with two parameters, $T$
and $p$. It is expected that for low $T$ and $p$ the system will be
magnetically ordered. In the ground states each sublattice has aligned
spins and thus the sublattice magnetization is a good order parameter:
\begin{equation}
	\label{eq:order_parameter}
	m = \frac{1}{N_\mathcal{P}}\sum_{i\in\mathcal{P}} s_i\,,
\end{equation}
where $N_{\mathcal P} = L^2/3$ ($L$ the linear system size) represents
the number of spins in one of the sublattices. The threshold $p_c$ for
topological subsystem codes is recovered as the critical $p$ along the
Nishimori line \cite{nishimori:81}
\begin{equation}
	\label{eq:nishimori}
	4\beta J = \ln\frac{1-p}{p/3}
\end{equation}
where the ferromagnetic phase of a sublattice is lost \cite{dennis:02}.

\paragraph*{Numerical details.---}
\label{sec:num}

\begin{figure}
\includegraphics[width=\columnwidth]{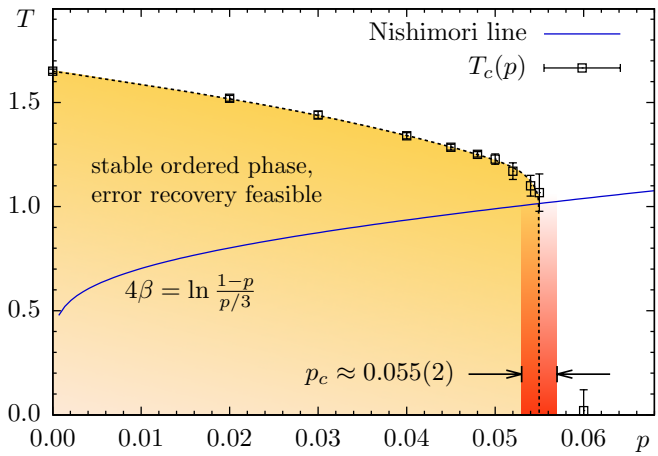}
\caption{(Color online)
Computed phase diagram for the classical disordered spin model shown in
Eq.~\eqref{eq:hamiltonian}. Each data point $T_c(p)$ on the phase
boundary (dashed curve separating white and shaded regions) is
calculated by locating the crossing in correlation function $\xi_L/L$
for different system sizes $L$ at a fixed disorder rate $p$.  The
Nishimori line (blue solid line) indicates where the requirement for the
mapping [Eq.~\eqref{eq:nishimori}] holds. The error threshold
$p_c\approx\theresult$ is found where the Nishimori line intersects the
phase boundary between the ordered phase (shaded) and the disordered
phase (not shaded, larger $T$ and $p$). Below $p_c\approx\theresult$
error correction is feasible. The (red) shaded vertical bar corresponds to
the statistical error estimate for $p_c$.
}
\label{fig:phasediagram}
\end{figure}

We investigate the critical behavior of the classical Ising spin model
[Eq.~\eqref{eq:hamiltonian2}] via large-scale parallel tempering Monte
Carlo simulations \cite{hukushima:96,katzgraber:06a}. Both spin states
and interaction terms are bit encoded to allow for efficient local
updates via bit masking. Detecting the transition temperature $T_c(p)$
for different fixed amounts of disorder allows us to pinpoint the phase
boundary in the $p$\,--\,$T$ phase diagram
(Fig.~\ref{fig:phasediagram}).

We choose periodic boundary conditions keeping in mind the colorability
requirements.  Then we can use the magnetization defined in
Eq.~\eqref{eq:order_parameter} to construct the wave-vector-dependent
magnetic susceptibility
\begin{equation}
\chi_m(\mathbf{k}) =
	\frac{1}{N_\mathcal{P}}\aT{\left(
	\sum_{i\in\mathcal{P}} S_i
	{\rm e}^{i\mathbf{k}\cdot\mathbf{R}_i}
	\right)^2} \, ,
\label{eq:susceptibility}
\end{equation}
where $\aT{\cdots}$ denotes a thermal average and $\mathbf{R}_i$ is the
spatial location of the spin $s_i$. From Eq.~\eqref{eq:susceptibility}
we construct the two-point finite-size correlation function,
\begin{equation}
\xi_L = \frac{1}{2\sin(k_{\rm min}/2)}\sqrt{
	\frac{\av{\chi_m(\mathbf{0})}}
	{\av{\chi_m(\mathbf{k}_{\rm min})}}-1
}\,,
\label{eq:correlation_function}
\end{equation}
where $\av{\cdots}$ denotes an average over disorder and
$\mathbf{k}_{\rm min} = (2\pi/L,0)$ is the smallest non-zero wave
vector.  Near the transition $\xi_L$ is expected to scale as
\begin{equation}
\xi_L/L \sim \tilde X[L^{1/\nu}(T-T_c)]\,,
\label{eq:xiL_scaling}
\end{equation}
where $\tilde X$ is a dimensionless scaling function. Because at the
transition temperature $T=T_c$, the argument of
Eq.~\eqref{eq:xiL_scaling} is zero (up to scaling corrections) and hence
independent of $L$, we expect lines of different system sizes to cross
at this point. If, however, the lines do not meet, we know that no
transition occurs in the studied temperature range.

\begin{figure}
\hfill\includegraphics[width=.977\columnwidth]{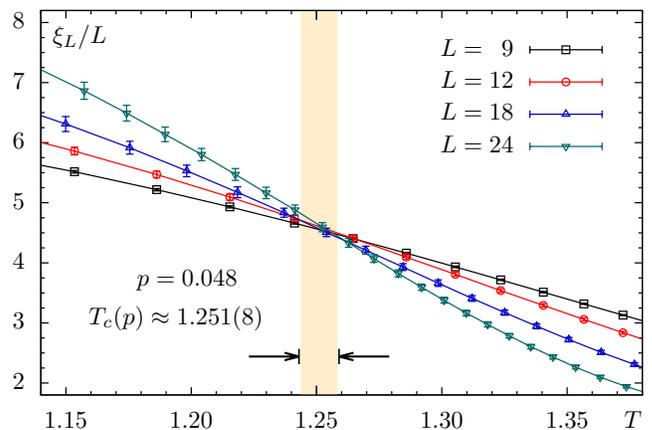}
\caption{(Color online)
Crossing of the correlation function $\xi_L/L$ with a disorder rate
of $p=\thesample$. The data exhibit a clear crossing at a transition
temperature of $T_c(p)\approx \thesampleTc$ \cite{comment:units}. The
shaded area corresponds to the error bar in the estimate of
$T_c(p)$. Note that error bars are calculated using a bootstrap
analysis of 500 resamplings. Corrections to scaling are minimal at
this disorder rate, but increase closer to the error threshold.
}
\label{fig:crossing}
\end{figure}

When determining the transition temperature $T_c(p)$ for a given
disorder rate $p$, the correlation functions $\xi_L/L$ are obtained
by averaging over several disorder realizations (governed by $p$)
for every system size $L$. Because we are only able to investigate
limited system sizes $L<\infty$, a careful analysis of finite-size
effects is required when estimating the transition temperature in
the thermodynamic limit.

In all simulations, equilibration is tested using a base-2 logarithmic
binning of the data: Once the data for all observables agree for three
logarithmically sized bins within error bars we deem the Monte Carlo
simulation for that system size to be in thermal equilibrium. The
simulation parameters can be found in Table~\ref{tab:simparams}.

\begin{table}[!tb]
\caption{
Simulation parameters: $p$ is the error rate for the depolarizing
channel, $L$ is the linear system size, $N_{\rm sa}$ is the number of
disorder samples, $t_{\rm eq} = 2^{b}$ is the number of equilibration
sweeps, $T_{\rm min}$ [$T_{\rm max}$] is the lowest [highest]
temperature, and $N_{\rm T}$ the number of temperatures used.
}
\label{tab:simparams}
\vspace*{2mm}
\centering
{\footnotesize
\begin{tabular*}{8 cm}{@{\extracolsep{\fill}} l r r r r r r}
\hline\hline
$p$ & $L$ & $N_{\rm sa}$ & $b$ & $T_{\rm min}$ & $T_{\rm max}$ &$N_{\rm T}$\\
\hline
$0.000$ -- $0.020$ & $9,12$  & $3\,200$  & $17$ & $1.40$ & $2.50$ & $24$\\
$0.000$ -- $0.020$ & $18$    & $1\,600$  & $18$ & $1.40$ & $2.50$ & $24$\\
$0.000$ -- $0.020$ & $24$    & $400$     & $19$ & $1.40$ & $2.50$ & $28$\\
$0.030$ -- $0.040$ & $9,12$  & $4\,800$  & $18$ & $1.25$ & $2.40$ & $28$\\
$0.030$ -- $0.040$ & $18$    & $2\,400$  & $19$ & $1.25$ & $2.40$ & $28$\\
$0.030$ -- $0.040$ & $24$    & $800$     & $20$ & $1.25$ & $2.40$ & $32$\\
$0.045$ -- $0.060$ & $9,12$  & $9\,600$  & $19$ & $0.9$ & $2.20$ & $32$\\
$0.045$ -- $0.060$ & $18$    & $4\,800$  & $21$ & $0.9$ & $2.20$ & $36$\\
$0.045$ -- $0.060$ & $24$    & $2\,400$  & $24$ & $0.9$ & $2.20$ & $48$\\
\hline\hline
\end{tabular*}}
\end{table}

\paragraph*{Results.---}
\label{sec:res}

For the pure system ($p=0$) there is a sharp transition visible directly
in the sublattice magnetization. The transition temperature $T_{c,{\rm
pure}}\approx 1.65(1)$ has not been computed before. For
larger amounts of disorder, a possible transition can be located
precisely by means of the two-point finite-size correlation function
[Eq.~\eqref{eq:correlation_function}]. Sample data for a disorder
strength of $p=\thesample$ (i.e., this would mean that on average
$\thesamplepercent$ of the physical qubits have failed) are shown in
Fig.~\ref{fig:crossing}, indicating a transition temperature of
$T_c(p)=\thesampleTc$. At $p=\theresult$, the lines only touch
marginally such that both the scenario of a crossing as well as no
transition are compatible within error bars. For error rates $p>p_c$,
the lines do not meet, indicating that there is no transition in the
temperature range studied.

The crossing of the critical phase boundary $T_c(p)$ with the Nishimori
line [Eq.~\eqref{eq:nishimori}] determines the error threshold to
depolarization. Our (conservative) estimate is $p_c\approx\theresult$.
Our results are summarized in Fig.~\ref{fig:phasediagram}, which shows
the estimated phase diagram.

\paragraph*{Summary.---}
\label{sec:sum}

We have calculated numerically the error resilience of topological
subsystem codes to the depolarizing channel by mapping the error
correction procedure onto a statistical-mechanical Ising spin model with
disorder.  The large critical error rate of $p_c = \theresultpercent$,
combined with a streamlined error recovery procedure that requires only
two-qubit interactions, constitutes a promising implementation concept
for quantum computing.

\paragraph*{Acknowledgments.---}

M.A.M.-D.~and H.B.~thank the Spanish MICINN Grant No.~FIS2009-10061, CAM
research consortium QUITEMAD S2009-ESP-1594, European Commission PICC:
FP7 2007-2013, Grant No.~249958, and UCM-BS Grant No.~GICC-910758. Work at
the Perimeter Institute is supported by Industry Canada and Ontario MRI.
H.G.K.~acknowledges support from the SNF (Grant No.~PP002-114713) and
the NSF (Grant No.~DMR-1151387). We thank ETH Zurich for CPU time on the
Brutus cluster and the Centro de Supercomputaci{\'o}n y
Visualisaci{\'o}n de Madrid (CeSViMa) for access to the Magerit-2
cluster.

\bibliography{refs,comments}

\begin{thebibliography}{33}
\bibitem[{com()}]{comment:units}
\bibinfo{note}{In the simulations we set the energy scale $J = 1$, i.e., all
  temperatures and estimates of $T_c$ are dimensionless.}
\end{thebibliography}

\end{document}